\documentclass[aip,jcp,reprint,citeautoscript,showpacs,floatfix]{revtex4-1}

\usepackage{amssymb}
\usepackage{amsmath}
\usepackage{amsfonts}
\usepackage{graphicx}

\newcommand{\fancy}{\mathcal}

\newcommand{\lbrs}{\left[}
\newcommand{\rbrs}{\right]}
\newcommand{\lbrc}{\left\{}
\newcommand{\rbrc}{\right\}}

\renewcommand{\vec}[1]{\boldsymbol{#1}}

\newcommand{\beq}{\begin{eqnarray}}
\newcommand{\eeq}{\end{eqnarray}}
\renewcommand{\d}{{\rm{d}}}
\newcommand{\tr}{{{\rm{Tr}}}}
\newcommand{\Tr}[1]{{{\rm{Tr}}\lbrs #1 \rbrs}}

\newcommand{\half}{\frac{1}{2}}
\newcommand{\aeq}{\approx}

\newcommand{\xrm}{\rm{x}}
\newcommand{\crm}{\rm{c}}
\newcommand{\xc}{\rm{xc}}
\newcommand{\Hx}{\rm{Hx}}

\newcommand{\Ec}{E^{\crm}}

\newcommand{\EHx}{E^{\Hx}}

\newcommand{\fx}{f^{\xrm}}
\newcommand{\fxc}{f^{\xc}}

\newcommand{\pr}{^{\prime}}

\newcommand{\vr}{\vec{r}}

\newcommand{\vrp}{\vec{r}\pr}

\newcommand{\vx}{\vec{x}}
\newcommand{\vxp}{\vec{x}\pr}

\newcommand{\vq}{\vec{q}}
\newcommand{\vqp}{\vec{q}\pr}

\newcommand{\FS}{\fancy{S}}

\newcommand{\KS}{{\rm{KS}}}
\newcommand{\Ext}{{\rm{Ext}}}

\newcommand{\vdW}{\rm{vdW}}

\newcommand{\chih}{\hat{\chi}}

\newcommand{\vh}{\hat{v}}
\newcommand{\wh}{\hat{w}}

\begin{document}
%\title{An effective and efficient kernel for ACFD %
%correlation energy calculations: the Radial Exchange Hole approximation}
%\title{Fixing correlation energy errors in the ACFD: %
%efficient screening via the ``Radial Exchange Hole'' kernel}
\title{Beyond the RPA on the cheap: improved correlation energies
with the efficient ``Radial Exchange Hole'' kernel}
\author{Tim Gould}\affiliation{Qld Micro- and Nanotechnology Centre, %
Griffith University, Nathan, Qld 4111, Australia}
\begin{abstract}
The ``ACFD-RPA'' correlation energy functional has been
widely applied to a variety of systems to successfully predict
energy differences, and less successfully predict absolute
correlation energies. Here we present a parameter-free
exchange-correlation kernel that systematically improves
absolute correlation energies, while maintaining most of the good
numerical properties that make the ACFD-RPA numerically tractable.
The ``RXH'' kernel is constructed to approximate the true exchange
kernel via a carefully weighted, easily computable radial averaging.
Correlation energy errors of atoms with two to eighteen electrons show a 
thirteenfold improvement over the RPA and a threefold
improvement over the related ``PGG'' kernel,
for a mean absolute error of 13mHa or 5\%.
The average error is small compared to all but the
most difficult to evaluate kernels. van der Waals $C_6$ coefficients
are less well predicted, but still show improvements on the RPA,
especially for highly polarisable Li and Na.
\end{abstract}
\maketitle
%Petersilka,Gossmann, Gross (PGG) approximation [PRL {\bf 76}, 1212 (1996)]

Standard groundstate DFT methods\cite{HohenbergKohn,*KohnSham}
under local density functional approaches such as the LDA, GGA
or exact exchange (EXX)\cite{OEP1,*OEP2} approach have long been
a fundamental part of computational physics and chemistry.
However, these functionals fail to reproduce important van der Waals (vdW)
dispersion physics\cite{Dobson2012-JPCM}.
%, which must be reintroduced via semi-empirical
%methods\cite{Dion2004,*Langreth2005,Vydrov2009,Tkatchenko2009}.
With increasing computer resources, the ACFD-RPA
functional approach, recently reviewed in
Eshuis, Bates and Furche\cite{Eshuis2012} (EBF) and
Ren~\emph{et al}\cite{Ren2012},
has proved to be a versatile and fairly
efficient\cite{Dobson1999,Furche2001,Miyake2002,Fuchs2002,%
Harl2008,*Harl2009,*Harl2010,Lebegue2010,Schimka2010}
means of evaluating correlation energies, albeit one that
typically overpredicts absolute correlation energies due to
its neglect of physics found in the exchange-correlation
kernel $\fxc$.

The ACFD-RPA is typically employed as a
``post groundstate'' method, where a groundstate Kohn-Sham (KS)
potential\cite{KohnSham} is generated self-consistently
in a standard approach (eg. the LDA or EXX) and the orbitals are
then used to refine the exchange-correlation energy without
further self-consistency. Here the bare response function
$\chi_0(\vx,\vxp;i\omega)$ is evaluated e.g. from the occupied
and unoccupied orbitals of a groundstate KS system and used to evaluate
the interacting response via
\begin{align}
\chi_{\lambda}=&\chi_0+\chi_0\star w_{\lambda}\star\chi_{\lambda}
\end{align}
where $\star$ represents an integral/sum over $\vx\equiv \vr\sigma$.
The effective interaction term $w_{\lambda}$ takes the form
\begin{align}
w_{\lambda}(\vx,\vx;i\omega)\equiv &v^C_{\lambda}(|\vr-\vrp|)
+\fxc_{\lambda}(\vx,\vxp;i\omega)
\label{eqn:wlambda}
\end{align}
where $v^C_{\lambda}(R)=\lambda/R$ is a Coulomb potential and
$\fxc_{\lambda}$ is the exchange-correlation kernel. Often
$\fxc_{\lambda}(\vx,\vxp;i\omega)$ is replaced by its
zero frequency expression
\begin{align}
\fxc_{\lambda}(\vx,\vxp;0)
=&\frac{\delta V^{\xc}_{\lambda}(\vx)}{\delta n(\vxp)}
\equiv \frac{\delta^2 E^{\xc}_{\lambda}}{\delta n(\vx)\delta n(\vx)},
\end{align}
or set to zero under the random-phase approximation (RPA).
Finally the correlation energy can be evaluated through
the so-called ``ACFD'' functional 
(see Section 2.1 of EBF\cite{Eshuis2012} for details)
\begin{align}
\Ec=&-\half\int_0^1\d\lambda\int_0^{\infty}\frac{\d \omega}{\pi}
\Tr{(\chi_{\lambda}-\chi_0)\star v^C_1}.
\label{eqn:Ec1}
\end{align}
This is then combined with the KS kinetic energy $T^{\KS}$,
external energy $E^{\Ext}$ and Hartree and exchange energy $\EHx$ to
produce the electronic groundstate energy
\begin{align}
E_0=&T^{\KS}+E^{\Ext}+\EHx+\Ec.
\end{align}

By neglecting the exchange-correlation kernel $\fxc$, the
RPA has some considerable numerical advantages that are often
exploited [see eg. Section 3 of EBF\cite{Eshuis2012},
Ref.~\onlinecite{Gruneis2009} and Ref.~\onlinecite{Ren2012}]
to make calculations numerically efficient. However, while
the ACFD-RPA often predicts good energy differences, it is
quite poor at predicting absolute correlation energies,
especially in small atoms and molecules\cite{Jiang2007} where it
can overpredict correlation energies by as much as 150\%.
These inaccuracies come from failures in the short-range
physics of interactions, arising from neglect of the $\fxc$ kernel.
For example the opposite-spin pair-density $n_{2\sigma\neq\sigma'}(\vr,\vrp)$
arising from the RPA can be negative for $\vr\aeq \vrp$,
a physically implausbile result that in turn contributes
to overestimation of the absolute correlation energy. When taking
energy differences these short-range errors tend to
cancel each other out, which explains the RPA's success for
these calculations.

Various approximations to $\fxc$, and other means to correct
RPA absolute energies have been proposed and can be broadly divided
into additive methods\cite{Kurth1999,Jiang2007} that
do not much affect dispersion interactions\cite{Dobson2012-JPCM}
and kernel methods
\cite{ALDA,Dobson2000,Savin1995,*Toulouse2004,*Toulouse2010,%
RPAx,PGG,Gruneis2009,Hellgren2008,Hesselmann2010,%
ISTLS,*Gould2012-2,Olsen2012} that do.
The most successful (and most numerically difficult) for atoms
and molecules is the
tdEXX\cite{Hellgren2008,Hesselmann2010} functional. Here the
exact exchange kernel $\fx$ is used to produce very accurate
correlation energies, suggesting that the exchange kernel is
sufficient for all but the most intractable problems. Indeed many
techniques\cite{RPAx,PGG,Jiang2007,Gruneis2009,Hellgren2008,Hesselmann2010}
base their corrections on exchange alone.

Much of the numerical advantage of the RPA comes from the fact
that $w_{\lambda}$ is: i) linear in $\lambda$, ii) frequency independent,
and iii) a function of $|\vr-\vrp|$ only. This allows one to perform
the integration over $\lambda$ analytically, the integration
over frequency efficiently, and, perhaps most importantly,
reduce the complexity of convolutions ($\star$) involving
$w_{\lambda}$ by transforming to reciprocal space.
For kernels with these three properties we can write the correlation
energy as (in Hartree atomic units, used throughout this manuscript)
\begin{align}
\Ec=&-\int_0^{\infty}\frac{\d\omega}{\pi}
\tr\lbrs \lbrc \log(\hat{1}+\chih_0\wh_1)
-\chih_0\wh_1\rbrc \frac{\vh_1}{\wh_1}
\rbrs
\label{eqn:EcDiag}
\end{align}
where $\chih_0(\vq,\vqp;i\omega)$ is a two-point function
in $\vq$ and $\vqp$ while $\wh_1(\vq,\vqp)=w_1(\vq)\delta(\vq-\vqp)$
and $\frac{\vh_1}{\wh_1}(\vq,\vqp)=\frac{v_1(\vq)}{w_1(\vq)}\delta(\vq-\vqp)$
are diagonal one-point functions. Clearly
$w_1(\vq)=v_1(\vq)=\frac{4\pi}{q^2}$ in the RPA, but
provided $\fxc_1$ is a function of $|\vr-\vrp|$ only,
both $\wh_1$ and $\vh_1/\wh_1$ are diagonal and \eqref{eqn:EcDiag} can
be efficiently evaluated. Most approximations to the true kernel
cannot be written in this form, making routine application
of kernel physics difficult. The ``Radial Exchange Hole'' (RXH)
kernel we shall develop in this manuscript aims to maintain these
good numerical properties, while reproducing some good
exchange kernel physics.

An exchange kernel of intermediate accuracy and complexity is that
proposed by Petersilka, Gossmann and Gross (PGG)\cite{PGG}.
They suggest approximating the exchange kernel $\fx$ via
the approach used by Krieger, Li and Iafrate\cite{KLI1992}
for the optimised exchange potential.
One can quickly `derive' their kernel by first writing the
Hartree and exchange energy as
$\EHx=\int\frac{\d\vx\d\vxp}{2|\vr-\vrp|}g_{\Hx}(\vx,\vxp)n(\vx)n(\vxp)$,
where $g_{\Hx}$ is the Hartree-exchange (Hx) pair-factor
defined via the groundstate Hartree and exchange pair-density $n_{\Hx}$
through $g_{\Hx}(\vx,\vxp)n(\vx)n(\vxp)=n_{\Hx}(\vx,\vxp)$.
By assuming $\delta g_{\Hx}/\delta n\aeq 0$ when evaluating
$\fxc_1(\vx,\vxp)=\delta^2 \EHx/[\delta n(\vx)\delta n(\vxp)]-v_1^C$,
one obtains the PGG kernel in the form
$\fxc_{\lambda}(\vx,\vxp)=\lambda[g_{\Hx}(\vx,\vxp)-1]/|\vr-\vrp|$
where $g_{\xrm}=[g_{\Hx}-1]<0$ is the exchange hole pair-factor.
This kernel has a number of good short-ranged properties
coming from the use of the exchange hole $g_{\xrm}$. Notably
it fully cancels the same spin Coulomb interaction for $\vr=\vrp$,
and re-introduces it quadratically for small $\vr-\vrp$. For
H and He the kernel reproduces the true $\fx=-\delta_{\sigma\sigma'}v^C$
and spurious self-interaction is completely avoided.

We now use an approximation similar to that of PGG to
produce the more efficient RXH kernel.
Here we aim to reproduce the good short-range physics of
the PGG kernel, while replacing the two-point behaviour of $g_{\Hx}$
by a simpler dependence on the distance $|\vr-\vrp|$ only.
We choose\footnote{The choice of function is quite arbitrary and
offers much scope for refinement.} a two parameter,
``radial'' model pair-factor $g$ of the form
\begin{align}
g(\vx,\vxp)\equiv g_{\sigma\sigma'}(R)=&
\frac{c_{\sigma\sigma'}R^2+(k_{\sigma\sigma'}R)^4}%
{1+(k_{\sigma\sigma'}R)^2+(k_{\sigma\sigma'}R)^4}
\label{eqn:gP}
\end{align}
where $R=|\vr-\vrp|$ and the two parameters control the on-top
curvature ($c$) and the range of the kernel ($k$).
Clearly $g(R\to 0)=c_{\sigma\sigma'} R^2$ and $g(R\to\infty)=1$
as desired (except for H and He where $c=k=0$ correctly makes
$g_{\sigma\sigma}=0$ everywhere). The total interaction
$w_{\lambda}=v^C_{\lambda}+\fxc_{\lambda}$ takes the desirable form
\begin{align}
w_{\lambda}(\vx,\vxp)=&\lambda\frac{g_{\sigma\sigma'}(|\vr-\vrp|)}{|\vr-\vrp|}.
\end{align}

We now introduce some ``exchange hole'' constraints to the
model kernel. Rather than allowing the parameters
$c_{\sigma\sigma'}$ and $k_{\sigma\sigma'}$ to be freely chosen, we
instead ensure that the kernel has some integrated properties
of the true pair-factor. Specifically we ensure that,
for each spin pair $\sigma\sigma'$, the Hx pair
number and groundstate Hx energy are reproduced via
\begin{align}
\int\d\vr\d\vrp[g(\vx,\vxp)n(\vx)n(\vxp)-n_{\Hx}(\vx,\vxp)]
=&0,
\label{eqn:NHxCond}
\\
\int\frac{\d\vr\d\vrp}{|\vr-\vrp|}
[g(\vx,\vxp)n(\vx)n(\vxp)-n_{\Hx}(\vx,\vxp)]
=&0.
\label{eqn:EHxCond}
\end{align}
The second constraint means that the `derivation' of the PGG kernel
given previously (where we assumed $\delta g/\delta n\aeq 0$ in
$\delta^2\EHx/[\delta n\delta n]$)
can be made for the RXH, since
$\EHx=\int\frac{\d\vx\d\vxp}{2|\vr-\vrp|}nn' g_{\Hx}
=\int\frac{\d\vx\d\vxp}{2|\vr-\vrp|}nn' g$. The kernel might
thus be considered a weighted radial averaging of the PGG kernel

The integrals give the model pair-factor $g$ some of the
same physics as the true pair-factor $g_{\Hx}$.
In particular they ensure that the dominant contribution from the
kernel is short-ranged, with long-ranged effects only being
felt in regions where the density (and thus response) is almost
zero. This in turn helps the RXH make its most important changes
to the correlation physics in regions with greatest contribution
to the energy. Clearly the RXH kernel behaves quite differently
to the the PGG kernel away from $\vr=\vrp$. For example, for
electrons in different $s$ shells (as in Li or Be) the PGG kernel
$g_{\Hx}$ depends only on $|\vr|$ and $|\vrp|$ and is exactly zero
whenever $|\vr|=|\vrp|$, while the RXH $g$ depends on
$|\vr-\vrp|$ and is non-zero for electrons with $|\vr|=|\vrp|$
but $\vr\neq\vrp$. However, if the effect of these differences
on the correlation energy act in similar regions to those
that dominate the exchange energy, the integral conditions
\eqref{eqn:NHxCond} and \eqref{eqn:EHxCond} should ensure
that the effect on the correlation energy is minimal.
In fact, as results later show, the RXH kernel appears to
better reproduce the true short-range physics of $\fxc$
than the PGG kernel, a somewhat surprising result.

To test the RXH kernel, we apply it to atoms with two to eighteen
electrons (He to Ar), using exact exchange groundstates
calculations. Here we evaluate the groundstate properties
under the LEXX\cite{Gould2012-LEXX} approximation to allow
for spherically symmeteric and spin homogeneous densities.
We use the same all-electron code previously employed to evaluate
ISTLS correlation energies\cite{Gould2012-2} with
groundstate optimised effective potential approximated
by the method of Krieger, Li and Iafrate\cite{KLI1992} (KLI).
The code is modified to reproduce the correct Hartree and
exchange physics of ensembles via the method outlined
in Ref.~\onlinecite{Gould2012-LEXX}. We are thus able to
reproduce groundstate properties, and in particular $n_{2\Hx}$,
accurately for both closed and open shell systems.

The linear-response calculation of $\chi_0$ involves utilising
expansions on the spherical harmonics to produce coupled
functions of radial coordinates $r=|\vr|$ and $r\pr=|\vrp|$ only.
To generate $\chi_0$ we avoid sums over unoccupied orbitals and instead
use Greens functions generated via a shooting method with
errors coming only from the finite radial grid with $N_r<400$ points.
Energies were converged to under 1mHa or within 0.5\% (whichever
is the larger).
Given the simple radial geometry we do not transform
integrals to reciprocal space, and the RPA, PGG and RXH methods
all take the same $O(N_r^3)$ time.

The parameters $c_{\sigma\sigma'}$ and $k_{\sigma\sigma'}$ of
the RXH kernel \eqref{eqn:gP}
are evaluated by employing the secant method. Here, for
a given $k$, we find $c$ via \eqref{eqn:NHxCond},
and $k$ is then iterated to satisfy \eqref{eqn:EHxCond}.
Typically no more than ten iterations are required to accurately
calculate $k$, although care must be taken to ensure convergence.
Some example parameters ($c_{\sigma\sigma}/k_{\sigma\sigma}$) are:
Be) $c=0.127$, $k=0.732$,
O) $c=0.667$, $k=2.635$, Na) $c=2.313$, $k=3.475$,
Si) $c=4.583$, $k=4.744$, and Ar) $c=11.241$, $k=5.692$.

\begin{table}[th]
\caption{Correlation energies in -mHa under the
RPA, PGG and RXH approximations. Numbers in brackets are \% errors.
Groundstates are calculated in the LEXX approximation.\label{tab:Ec}}
\begin{ruledtabular}
\begin{tabular}{lrrrrrrr}
System & Exact &   RPA &   \%  &   PGG &   \%  &   RXH &   \%  \\
\hline
   He  &    42 &    84 & (100) &    45 &   (7) &    45 &   (7) \\
   Li  &    45 &   113 & (151) &    50 &  (11) &    50 &  (11) \\
   Be  &    94 &   181 &  (93) &   104 &  (11) &   106 &  (12) \\
    B  &   125 &   224 &  (79) &   112 & (-10) &   128 &   (3) \\
    C  &   156 &   277 &  (78) &   124 & (-21) &   155 &  (-1) \\
    N  &   188 &   337 &  (79) &   139 & (-26) &   175 &  (-7) \\
    O  &   258 &   406 &  (57) &   206 & (-20) &   243 &  (-6) \\
    F  &   324 &   491 &  (52) &   270 & (-17) &   304 &  (-6) \\
   Ne  &   390 &   586 &  (50) &   332 & (-15) &   369 &  (-5) \\
   Na  &   396 &   611 &  (54) &   332 & (-16) &   379 &  (-4) \\
   Mg  &   438 &   674 &  (54) &   375 & (-14) &   436 &  (-0) \\
   Al  &   470 &   715 &  (52) &   385 & (-18) &   467 &  (-1) \\
   Si  &   505 &   772 &  (53) &   401 & (-21) &   506 &   (0) \\
    P  &   540 &   839 &  (55) &   419 & (-22) &   552 &   (2) \\
    S  &   605 &   903 &  (49) &   474 & (-22) &   621 &   (3) \\
   Cl  &   666 &   982 &  (47) &   527 & (-21) &   694 &   (4) \\
   Ar  &   722 &  1072 &  (48) &   579 & (-20) &   773 &   (7) \\
\hline
MAE(\%) &       &   194 &  (68) &    66 &  (17) &    13 &   (5) \\
\end{tabular}
\end{ruledtabular}
\\$^a$ From Ref.~\onlinecite{Chakravorty1993}
\end{table}
Absolute correlation energies $-\Ec$ for the atoms are shown
in Table~\ref{tab:Ec} under the RPA, PGG and RXH approximations.
It is immediately clear that, with the exception of Be,
the RXH outperforms not only the RPA as expected, but
also the PGG kernel. Here the mean absolute error of $\Ec$ is reduced
from 194mHa for the RPA and 66mHa for the PGG down to
13mHa for the RXH. Absolute percentage errors are reduced
somewhat less, but even here RXH offers a threefold improvement
from 17\% to 5\% over the PGG kernel. For atoms with
highest occupied orbitals in a $p$ shell, the improvement is
even greater still, with correlation energies approaching
chemical accuracy.

It is clear that the RXH must offer an unexpected improvement
over the PGG in the regions that dominate correlation energy
calculations. It is likely that the spherical averaged form
and integrated constraints correctly confine the kernel to be
shorter-ranged than the PGG kernel. The dynamic physics will
thus be reproduced better near the crucial ``on-top'' region.
In the case of the transitions from $p$ to $s$ that dominate
correlation energies for outermost $p$ shells the RXH kernel
significantly outperforms the PGG kernel, adding further weight
to the assumption that it is the short range physics that
are most important. It would be very useful to compare the
exact exchange kernel $\fx$ of Ne from a tdEXX
calculation\cite{Hellgren2008,Hesselmann2010} with
its RXH kernel.

\begin{table}[th]
\caption{Same-species $C_6$ coefficients (Ha$a_0^6$) for spherical
atoms in the RPA, PGG and RXH approximations.\label{tab:C6}}
\begin{ruledtabular}
\begin{tabular}{lrrrrrrr}
& He & Li & Be & Ne & Na & Mg & Ar
\\ \hline
Exact$^a$ & 1.44 & 1380 & 219 & 6.48 & 1470 & 630 & 63.6 \\
RPA   & 1.17 &  500 & 179 & 4.98 &  744 & 481 & 54.7 \\
PGG   & 1.38 & 1340 & 277 & 5.98 & 1710 & 743 & 76.0 \\ 
RXH   & 1.38 & 1080 & 231 & 5.19 & 1050 & 487 & 55.2 \\
\end{tabular}
\end{ruledtabular}
\\$^a$ From lower bounds of Ref.~\onlinecite{StandardCertain}
\end{table}
We have so far demonstrated that the RXH kernel is good at
reproducing the intra-species correlations that feed into
the correlation energy. To test its ability to predict
inter-species correlations we also use the RXH kernel
to evaluate van der Waals $C_6$ coefficients for same-species pairs.
These are notoriously difficult to reproduce accurately under
standard DFT or ACFD type approaches.
Here the correlation energy difference of two atoms at a great
distance $D$ obeys $E^{\vdW}(D)=\Ec(D)-\Ec(\infty)=-\frac{C_6}{D^6}$.
We calculate $C_6$ through the Casimir-Polder formula
\begin{align}
C_6=&\frac{3}{\pi}\int\d\omega A(i\omega)^2,
\end{align}
where $A(i\omega)=\int\d\vx\d\vxp zz'\chi_1(\vx,\vxp)$
is the dipole polarisibility of a single atom.

As shown in Table~\ref{tab:C6} the results are mixed, with the
RXH kernel being only slightly more accurate than the RPA for the large
Mg and Ar atoms, but otherwise outperforming it. Importantly
it greatly improves on the RPA for Li and Na, the systems with
the greatest, and thus most dominant vdW $C_6$ coefficients.
We note that the evaluation of $C_6$ coefficients is quite
susceptible to even minor numerical issues and small variations
in the groundstate orbitals, and these results may need to be
better converged and tested for a comprehensive understanding.

In summary, we have produced a ``Radial Exchange Hole''
(RXH) exchange-correlation kernel that
is both efficient to calculate, and systematically improves on
the RPA, at least for the atomic systems tested here. The
kernel is derived from a similar approximation to that of PGG
and is designed to model the exchange kernel $\fx(\vr,\vrp)$
via a simple dependence on $|\vr-\vrp|$ only.
By writing the kernel in the form \eqref{eqn:gP},
constrained by \eqref{eqn:NHxCond} and \eqref{eqn:EHxCond}
we are able to produce a kernel that is simple
and parameter free.

The RXH kernel offers improved correlation energies on all tested
atoms (He to Ar) when compared with both the RPA and related PGG kernel
(with the exception of Be). The mean absolute error of the atomic
correlation energies is reduced to 13mHa (5\%) from 66mHa (17\%)
and 194mHa (68\%) respectively for the PGG and RPA kernels.
Here the energy is dominated by the short-range physics correctly
modelled by the RXH kernel.
Results for vdW $C_6$ coefficients are more mixed, with predictions
only slightly better than the RPA for Mg and Ar, but significantly
better for Li, Be and Na. It is likely that longer-ranged kernel
physics are involved in these calculations, and these are more
poorly predicted by the RXH approximation.

While the RXH approximation derived and demonstrated here
clearly performs well for correlation energies of atoms, its
immediate application to molecules and atoms may prove
problematic. In particular the dissociation limit of different
atoms will involve finding a pair factor $g$ through \eqref{eqn:gP}
that is appropriate for both atoms. This problem could be
remedied by treating the kernel as a sum over localised
radial pair-factors
[e.g. $\fxc=\sum_{\FS}\Theta_{\FS}(\vr)\Theta_{\FS}(\vrp)
\frac{g_{\FS}(\vx,\vxp)-1}{|\vr-\vrp|}$ with $\Theta_{\FS}(\vr\notin\FS)=0$ for
given regions $\FS$],
which we will explore in future work.
Similarly in bulks, the form of the model \eqref{eqn:gP}
of the pair-factor is inappropriate, and an alternative
approximation must be found.

Overall, however, the RXH clearly represents an approximation
that is both numerically efficient, and more accurate than
the RPA. Somewhat surprisingly it produces more accurate
correlation energies than the related PGG approximation,
and many other approximations to $\fxc$.
It is clear that constraining the kernel
by its integrated properties allows it to accurately reproduce
the important short-ranged physics that the RPA neglects,
and this is likely to apply to other expressions for $g$.
It should thus be possible to further refine the form of the
kernel, perhaps via the addition of free parameters or
a similar approach to include the correlation kernel,
to improve both correlation energies and $C_6$ coefficients.

\acknowledgments
The author would like to especially thank John F. Dobson, and
Filipp Furche, Georg Jansen and Sebastien Leb\'egue, for much
fruitful discussion.
This work was supported by ARC Discovery Grant DP1096240.

%\section*{References}
\bibliography{vanDerWaals,ACFDT,DFT,Wannier,Misc,Experiment,Hybrid,%
QMGeneral,ISTLS,Frac}

\end{document}